\documentclass[aps,prb,preprint,groupedaddress,showkeys]{revtex4-2}
\usepackage{amstext}
\usepackage{amssymb}
\usepackage{amsmath,esint}
\usepackage{float}
\usepackage{graphicx}
\usepackage{color}
\usepackage{comment}
\newcommand{\comm}[1]{}
\usepackage{hyperref}
\usepackage{float}
\usepackage{bm}
\usepackage{enumerate}

\begin{document}


\title{Comment on "Effects of shear methods on shear strengths and deformation modes of two
	typical transition metal carbides and their unification"}


\author{Marcin Ma\'zdziarz}
\email[]{mmazdz@ippt.pan.pl}
\affiliation{Institute of Fundamental Technological Research Polish Academy of Sciences,
	Pawi\'nskiego 5B, 02-106 Warsaw, Poland}


\date{\today}

\begin{abstract}
{Recently, Chuanying Li, Tao Fu, Xule Li, Hao Hu, and Xianghe Peng in [\href{https://doi.org/10.1103/PhysRevB.107.224106}{Phys. Rev. B 107, 224106}] investigated the mechanical behavior of cubic HfC and TaC under simple shear (SS) and pure shear (PS) using first-principles calculations. Unfortunately, the paper contains some serious and fundamental flaws in the field of {continuum~mechanics} and nanomechanics. The results presented appear to be qualitatively and quantitatively incorrect, they would be correct if we were in the small/linear deformation/strain regime, which we are not. A correct description therefore requires a finite/nonlinear deformation/strain apparatus.}
\end{abstract}



\maketitle

\section{Introduction}
\label{sec:Int}

In the paper entitled "Effects of shear methods on shear strengths and deformation modes of two typical transition metal carbides and their unification" by Chuanying Li et al. \cite{Li2023} cubic HfC and TaC crystals were subjected to simple (SS) and pure shear (PS) by DFT simulations. The paper mistakes concepts from the continuum~mechanics, incorrectly enforces deformations, and it appears that the computational results are qualitatively and quantitatively incorrect.  














\begin{itemize}
	\item  The Authors write: {"During deformation, the lattice vectors need to be changed, ... Hence, the deformation can be imposed by transforming
the \textit{i-1}th step lattice vector matrix \textbf{R}$^{i-1}$ to the deformed \textit{i}th
step lattice vector matrix \textbf{R}$^i$ as follows [36]:}

\begin{equation}\label{eq1}
	\textbf{R}^i = \textbf{R}^{i-1}\left[\textbf{I}+\left( \begin{array}{ccc}
		{0} & {0} & {0} \\
		{0} & {0} & {0} \\ 
		{0} & {\Delta{\varepsilon_{zy}}} & {0}\\
	\end{array} \right)  \right] ."
\end{equation}

This formula is incorrect.

In the non-linear mechanics of crystals there is well know hypothesis, called the \emph{Cauchy-Born rule} \cite{BornHuang1954},  
which assumes that under a homogeneous macroscopic deformation, the primitive Bravais lattice vectors of a 3D crystal deform in an affine manner via a 3$\times$3 matrix \textbf{F}:
\begin{equation}
	\textbf{a}_i = \textbf{F} \textbf{A}_i,  
	\label{cb}
\end{equation}
where \textbf{a}$_i$ stands for a spatial lattice vectors, \textbf{A}$_i$ reference vectors and \textbf{F} the deformation gradient \cite{clayton2011nonlinear}.

If we even wanted to do it in steps, it is clear from the tensor calculus point of view that Eq.\ref{eq1} should be as follows:
\begin{equation}
	{{\bf R}^{def}} = {\bf F} \cdot {{\bf R}^{ini}}. \label{eq2}
\end{equation}
The vector, or matrix of vectors, is multiplied by the matrix on the left.

Moreover, the components of the deformation gradient \textbf{F} are not, in general, components of the strain tensor.
\end{itemize}

\begin{itemize}
	\item  
 The Authors write further: {"For PS, after each $\varepsilon_{zy}$, the atomic coordinates and the
other five independent strain components (except $\varepsilon_{zy}$) are optimized
simultaneously to reach a stress state with $\sigma_{xx}$ = $\sigma_{yy}$ =
$\sigma_{zz}$ = $\sigma_{xy}$ = $\sigma_{zx}$ = 0 [35,37,38]. For SS, after applying $\varepsilon_{zy}$,
the atomic coordinates are optimized but remain $\varepsilon_{xx}$ = $\varepsilon_{yy}$ =
$\varepsilon_{zz}$ = $\varepsilon_{xy}$ = $\varepsilon_{zx}$ = 0 [14,39]. The schematics of PS and SS are
illustrated in Figs. 1(a) and 1(b), respectively."}

The above excerpt from the Authors' text contains several misrepresentations. In general, the \emph{Lagrangian finite strain tensor} \textbf{E}, also called the \emph{Green-Lagrangian strain tensor} or \emph{Green – St-Venant strain tensor} is defined as: 
\begin{equation}\label{EGreen}
 {\bf E} = \frac{1}{2} ({\bf F}^T {\bf F} - {\bf I}) . 
\end{equation}

and its linear approximation the \emph{infinitesimal strain tensor} \textbf{{$\varepsilon$}}, also called the \emph{Cauchy's strain tensor}, \emph{linear strain tensor}, or \emph{small strain tensor} takes the form:
\begin{equation}\label{Esmall}
	{\bf \varepsilon} = \frac{1}{2} ({\bf F}^T +{\bf F}) - {\bf I} . 
\end{equation}

The classical \emph{finite simple shear deformation} (SS) is an isochoric plane deformation defined by the deformation gradient tensor \textbf{F} in the following form \cite{THIEL201957} (and this implies the form of the tensor {\bf E} and {\bf $\varepsilon$}):

\begin{equation}\label{fss}
	\textbf{F}^\text{SS} = \left[ \begin{array}{ccc}
		{1} & {0} & {0} \\
		{0} & {1} & {\gamma} \\ 
		{0} & {0} & {1}\\
	\end{array} \right] \Rightarrow 
    \textbf{E}^\text{SS} = \left[ \begin{array}{ccc}
	{0} & {0} & {0} \\
	{0} & {0} & {\frac{\gamma}{2}} \\ 
	{0} & {\frac{\gamma}{2}} & {\frac{{\gamma}^2}{2}}\\
    \end{array} \right] \Rightarrow 
    \mathbf{\varepsilon}^\text{SS} = \left[ \begin{array}{ccc}
	{0} & {0} & {0} \\
	{0} & {0} & {\frac{\gamma}{2}} \\ 
	{0} & {\frac{\gamma}{2}} & {0}\\
   \end{array} \right].
\end{equation}

It can be seen here that $\textbf{E}^\text{SS}_\text{zz}\neq 0$. This would be the case if we were using the small strain tensor  $\mathbf{\varepsilon}^\text{SS}_\text{zz}$. The deformations in this work are not small. It is important to remember what the consequences of overusing linear theories can be. Because \textbf{F} is not objective and hence small strain tensor \textbf{$\varepsilon$} also and a rigid rotation can induce any stresses in a deformable body and they should not, see \cite{Mazdziarz2019} . 

The \emph{pure shear deformation} (PS) \cite{ogden2013non} is a deformation in which the body is elongated in one direction while being shortened perpendicularly and is defined by:

\begin{equation}\label{fps}
	\textbf{F}^\text{PS} = \left[ \begin{array}{ccc}
		{1} & {0} & {0} \\
		{0} & {\beta} & {0} \\ 
		{0} & {0} & {\frac{1}{\beta}}\\
	\end{array} \right] \Rightarrow 
	\textbf{E}^\text{PS} = \left[ \begin{array}{ccc}
		{0} & {0} & {0} \\
		{0} & {\frac{{{\beta}^2-1}}{2}} & {0} \\ 
		{0} & {0} & {\frac{{{\beta}^{-2}-1}}{2}}\\
	\end{array} \right] \Rightarrow 
	\mathbf{\varepsilon}^\text{PS} = \left[ \begin{array}{ccc}
		{0} & {0} & {0} \\
		{0} & {\beta-1} & {0} \\ 
		{0} & {0} & {{\beta}^{-1}-1}\\
	\end{array} \right].
\end{equation}

The details of the PS in the rotated coordinate frame along the shear plane can be found in \cite{THIEL201957}. The Authors' PS is not in fact a PS, but another deformation.

The \emph{pure shear stress} (PSS) is such a deformation for which the Cauchy stress tensor is a pure shear stress tensor of the form \textbf{$\sigma$} = \emph{$\tau$}($e_2$ $\otimes$ $e_3$ +
$e_3$ $\otimes$ $e_2$) with \emph{$\tau$} $\in$ $\mathbb{R}$ \cite{THIEL201957}. To enforce such a deformation \textbf{F} must take the form
of a simple shear composed with a triaxial stretch:
\begin{equation}\label{fpss}
	\mathbf{\sigma}^\text{PSS} = \left[ \begin{array}{ccc}
		{0} & {0} & {0} \\
		{0} & {0} & {\tau} \\ 
		{0} & {\tau} & {0}\\
	\end{array} \right] \implies
	\textbf{F}^\text{PSS} = \left[ \begin{array}{ccc}
	{a} & {0} & {0} \\
	{0} & {b} & {c {\eta}} \\ 
	{0} & {0} & {c}\\
\end{array} \right] \Rightarrow 
\textbf{E}^\text{PSS} = \left[ \begin{array}{ccc}
	{\frac{a^2-1}{2}} & {0} & {0} \\
	{0} & {\frac{b^2-1}{2}} & {\frac{b c \eta}{2}} \\ 
	{0} & {\frac{b c \eta}{2}} & {\frac{{\eta}^2 c^2+c^2-1}{2}}\\
\end{array} \right] .
\end{equation}
\end{itemize}

\begin{itemize}
	\item 
	
In	"FIG. 2. Mechanical response of samples sheared along [1$\bar{1}$0](110). (a), (b) Variations of stress components of HfC and TaC with $\varepsilon_{zy}$ under SS and PS, respectively. ..." the Authors presented the results of their DFT shear computations.

The HfC and TaC crystals analyzed here have cubic symmetry and there is no reason for them to behave differently in shear than other cubic crystals, see \cite{Ogata_2009}. The shear component of the stress during SS deformation, by the way for PSS also, should have a \emph{sine}-like character. In order to verify this, the HfC[1$\bar{1}$0](110) crystal was subjected to SS and PSS by \emph{ab-initio} calculations based on density functional theory (DFT) \cite{DFT-HK, DFT-KS} within the pseudopotential plane-wave approximation (PP-PW) performed by the ABINIT \cite{GONZE2020107042} code. The projector augmented wave method (PAW) pseudopotentials used for the PBE \cite{Perdew1996} generalized gradient approximation (GGA) exchange–correlation functionals (XC) were obtained from the PseudoDojo project \cite{JOLLET20141246}. The valence electron configurations and the \emph{cut-off} energy used for Hf and C atoms were consistent with those utilized in the PAW pseudopotentials. The calculation accuracy settings correspond to those used in \cite{Mazdziarz2022}.

\begin{figure}[H] 
	\label{fig:HfC}
	\centering
	\begin{tabular}{ll}
		{\footnotesize \bf a)} & {\footnotesize \bf b)}\\
		\includegraphics[width=0.470\linewidth]{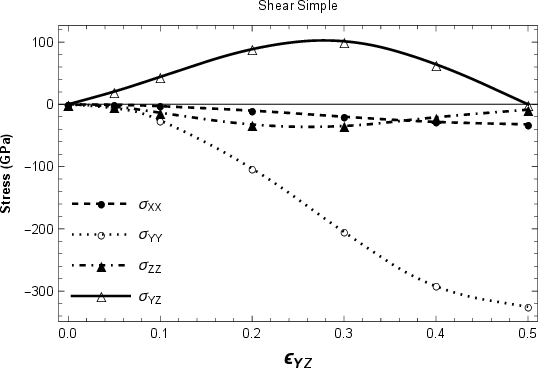} &
		\includegraphics[width=0.470\linewidth]{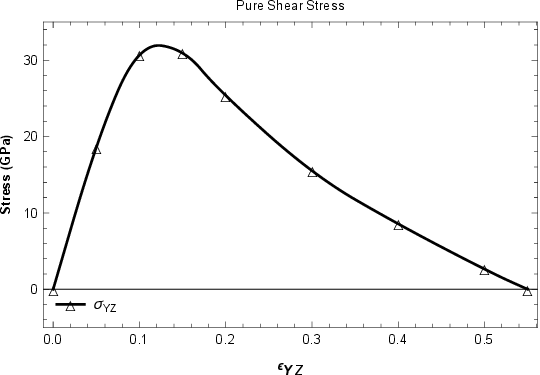} \\
	\end{tabular}
	\caption{HfC [1$\bar{1}$0](110): a) Simple Shear, b) Pure Shear Stress}
\end{figure}
\end{itemize}

It can be seen from Figure \ref{fig:HfC} that the results presented here are substantially different from those presented by the Authors in their Figure 2 for HfC. For SS, the $\sigma_{yz}$ shear component of the stress has a proper \emph{sine}-like character and the $\sigma_{yy}$ component of the stress is about twice as large. Zeroing the normal stresses in the PSS reduces the shear stress and introduces its asymmetry. 
These results also differ substantially from those presented by the Authors.

\begin{acknowledgments}
This work was partially supported by the National Science Centre (NCN -- Poland) Research Projects: No. 2021/43/B/ST8/03207. Additional assistance was granted through the computing cluster GRAFEN at Biocentrum Ochota, the Interdisciplinary Centre for Mathematical and Computational Modelling of Warsaw University (ICM UW) and Pozna\'n Supercomputing and Networking Center (PSNC).
\end{acknowledgments}



\bibliography{PRB}

\end{document}